\documentclass[onecolumn]{aastex6}

\usepackage{color,txfonts}
\usepackage{hyperref}
\usepackage{epstopdf}
\usepackage{graphicx}
\usepackage[FIGTOPCAP]{subfigure}

\newcommand{\yzf}[1]{{\color{red} #1}}

\begin{document}

\title{GRB 111005A at $z=0.0133$ and the prospect of establishing long-short GRB/GW association}

\author{Yuan-Zhu Wang$^{1,2}$, Yong-Jia Huang$^{1,3}$, Yun-Feng Liang$^{1}$, Xiang Li$^{1}$, Zhi-Ping Jin$^{1,3}$, Fu-Wen Zhang$^{4}$,
 Yuan-Chuan Zou$^{5}$, Yi-Zhong Fan$^{1,3}$, and Da-Ming Wei$^{1,3}$}
\affil{
$^1$ {Key Laboratory of dark Matter and Space Astronomy, Purple Mountain Observatory, Chinese Academy of Science, Nanjing, 210008, China.}\\
$^2$ {University of Chinese Academy of Sciences, Yuquan Road 19, Beijing, 100049, China}\\
$^3$ {School of Astronomy and Space Science, University of Science and Technology of China, Hefei, Anhui 230026, China.}\\
$^4$ {College of Science, Guilin University of Technology, Guilin 541004, China.}\\
$^5$ {School of Physics, Huazhong University of Science and Technology, Wuhan 430074, China.}\\
}
\email{xiangli@pmo.ac.cn (XL), jin@pmo.ac.cn (ZPJ), yzfan@pmo.ac.cn (YZF)}

\begin{abstract}
GRB 111005A, one long duration gamma-ray burst (GRB) occurred within a metal-rich environment that lacks massive stars with $M_{\rm ZAMS}\geq 15M_\odot$, is not coincident with supernova emission down to stringent limit and thus should be classified as a ``long-short" GRB (lsGRB; also known as SN-less long GRB or hybrid GRB), like GRB 060505 and GRB 060614. In this work we show that in the neutron star merger model, the non-detection of the optical/infrared emission of GRB 111005A requires a sub-relativistic neutron-rich ejecta with the mass of $\leq 0.01~M_\odot$, (significantly) less massive than that of GRB 130603B, GRB 060614, GRB 050709 and GRB 170817A. The lsGRBs are found to have a high rate density and the neutron star merger origin model can be unambiguously tested by the joint observations of the second generation gravitational wave (GW) detectors and the full-sky gamma-ray monitors such as Fermi-GBM and the proposing GECAM. If no lsGRB/GW association is observed in 2020s, alternative scenarios have to be systematically investigated. With the detailed environmental information achievable for the very-nearby events, a novel kind of merger or explosion origin may be identified.
\end{abstract}

\keywords{Gamma-ray burst, Gravitational Wave, Macronova, GRB 111005A}

\section{Introduction} \label{sec:intro}
Based on the duration of their prompt emission, the gamma-ray bursts (GRBs) are usually classified into two groups divided by $\sim 2$ seconds. The GRBs with a duration longer than 2 seconds are called as the long GRBs while the events with a shorter duration are called the short GRBs \citep{Kouveliotou1993,Piran2004}. Long GRBs are believed to
originate from collapsars that involve death of massive stars
and are expected to be accompanied by luminous supernovae
 \citep[SNe, see][]{Woosley2006} while the short GRBs are suspected to be from neutron star mergers and hence should not be coincident with luminous SNe \citep{Eichler1989}. Instead the short GRBs are likely associated with
Li-Paczy'{n}ski macronova (also called kilonova)$-$a novel kind of near-infrared/optical transient powered by the radioactive decay of
heavy elements synthesized in the ejecta of a compact binary merger \citep[e.g.,][]{1998ApJ...507L..59L,2013ApJ...774...25K,2013ApJ...775..113T,2017LRR....20....3M}. The collapsar origin of most long GRBs has been confirmed by the SN detection in the afterglow followup observations \citep{Woosley2006} while the neutron star merger model of short GRBs is supported by the observations of their
afterglows and host galaxies \citep{Berger2014} as well as the detection of macronovae in GRB 130603B \citep{Berger2013,Tanvir2013}, GRB 050709\citep{Jin2016} and GRB 170817A \citep[an event released after the submission of this work, see e.g.][]{von Kienlin2017,Pian2017,Covino2017}. Though the long and short classification has been widely adopted by the community, a few events, including GRB 060505 and GRB 060614 \citep{Fynbo2006}, share some characters of both groups (i.e., the so-called long-short GRBs or SN-less long GRBs, the events with apparent long duration but without luminous SN emission) and thus challenge the above simple scheme. The physical origin of these ``outliers" attracted wide attention and were widely debated in the literature. The theoretical studies have shown that $\sim 40\%$ of {\it Swift} bursts shorter
than 2 sec may in fact be from collapsars, and alternatively, a non-negligible amount
of non collapsars may have durations longer than 10 sec \citep[e.g.][]{Bromberg2013}. The long-short GRBs may be such non collapsars.
The identification of a macronova signature in the late afterglow of GRB 060614 \citep{Yang2015,Jin2015} provides a direct support to the hypothesis that (some) long-short GRBs (lsGRBs) are intrinsically ``short" \citep{Gehrels2006,Gal-Yam2006,ZhangB2007,Ofek2007}. For GRB 060505, the situation is however less clear and a novel massive star explosion origin is possible \citep{Fynbo2006,DellaValle2006,Xu2009} since the properties of its host galaxy are consistent with those
expected for canonical long-duration GRBs \citep{Thone2008}. In view of these uncertainties, more reliable probe of the nature of the progenitor stars of lsGRBs is crucially needed.
The presence of a lsGRB at a very low redshift $z=0.0133$ (i.e., GRB 111005A), as revealed very recently \citep{Michal2017,Tanga2017}, makes such a topic far more attractive than before.

GRB 111005A triggered the {\it Swift} BAT at 08:05:14 UT on 2011 Oct 5 \citep{Saxton2011}.
This burst has a duration of $T_{90}=26\pm7$ sec. Its fluence in the $15-150$ keV band is of $(6.2\pm1.1)\times
10^{-7}~{\rm erg~cm^{-2}}$ and the spectrum is best-fitted
by a single power-law with index $\Gamma=2.4\pm 0.2$ \citep{Tanga2017}. Therefore, GRB 111005A is likely a typical long-soft GRB.
 Due to sun site constraint, no X-ray or optical quick followup observations
were carried out. Near-infrared images
taken during twilight and close to the horizon did not reveal
any variable source, ruling out the presence of any bright SN emission \citep{Michal2017}. However, the followup radio observations located GRB111005A very accurately and thus established the association of GRB 111005A and the galaxy ESO 580-49 \citep{Michal2017}.
With such a low redshift, GRB 111005A is the second closest long GRB ever detected, making it the closet lsGRB and enabling the nearby environment to be studied at an
unprecedented resolution of $100~{\rm pc^{2}}$. From the analysis of the MUSE data cube, \citet{Tanga2017} found GRB 111005A to have occurred within a
metal-rich environment with little signs of ongoing star formation. Their spectral analysis at the position of the GRB indicates the presence
of an old stellar population ($\geq$10 Myr), which limits the mass of the GRB progenitor to $M_{\rm ZAMS}<15 M_\odot$, in direct conflict with the
collapsar model. The deep limits on the presence of any SN emission combined with the environmental conditions at the position of
GRB 111005A thus favor the non-collapsar origin (i.e., the merger origin). Among the possible non-collapsar scenarios,
the neutron star merger possibility is very attractive since the detection of one such a very-nearby event by {\it Swift} in about 10 years performance
points to a high rate and the second generation GW detectors, for example, the advanced LIGO/Virgo \citep{LVC2010}, can catch the signals.
The main purpose of this work is to examine whether the binary neutron star merger scenario really meets the infrared/optical upper limits and evaluate the prospect of establishing or ruling out the neutron star merger origin of lsGRBs in the advanced LIGO/Virgo era.

\section{Limits on the macronova emission associated with GRB 111005A}\label{sec:macronova-limits}
\begin{figure}[ht!]
\figurenum{1}\label{fig:comparision}
\centering
\includegraphics[angle=0,scale=0.6]{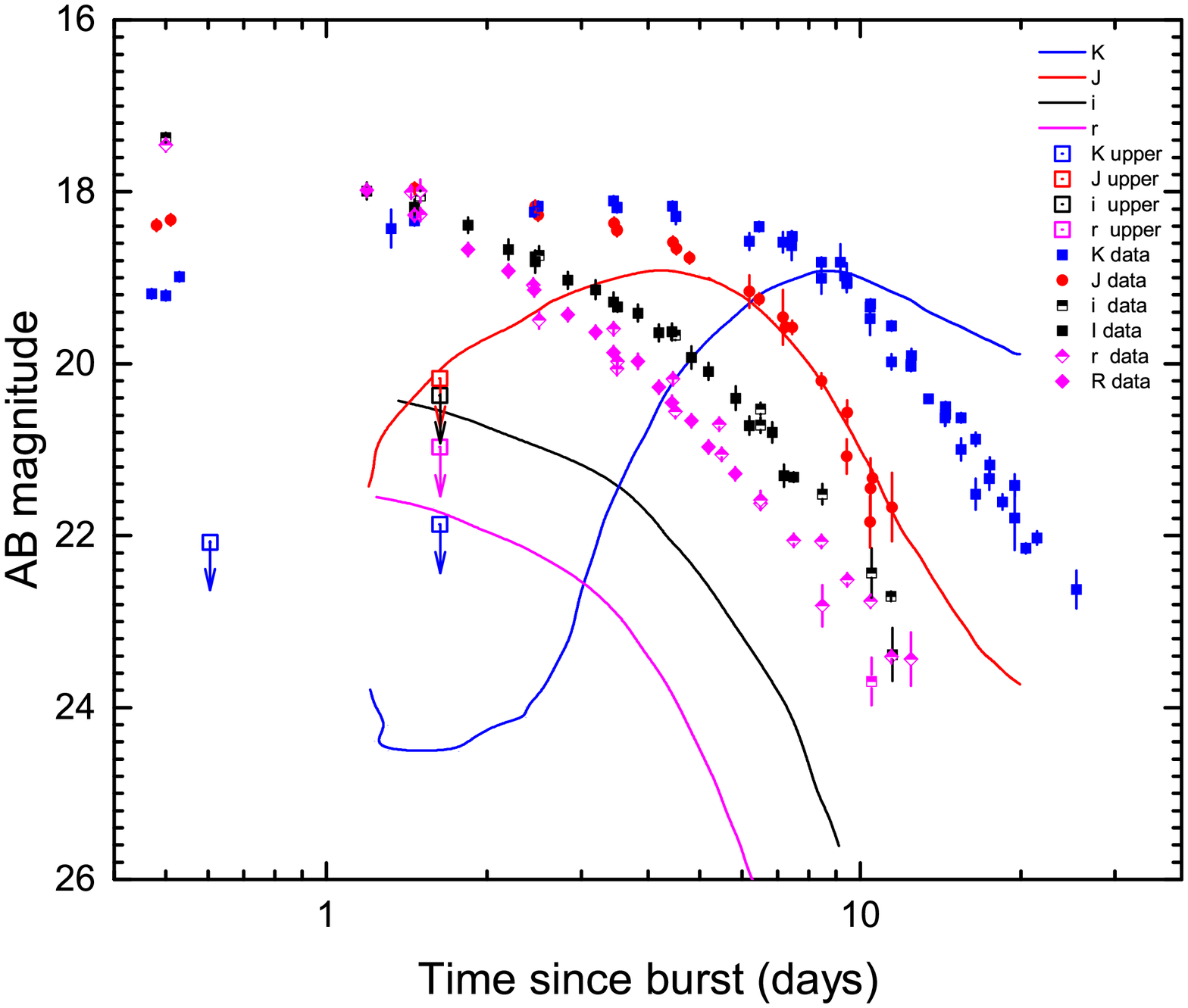}
\caption{The $r, ~i, ~J, ~K$ band upper limits of GRB 111005A \citep[adopted from][]{Michal2017} versus the macronova emission predicted in the NSM-all model of \citet[][shifted to $z=0.0133$]{2013ApJ...775..113T}, in which $M_{\rm ej}=0.01M_\odot$ and the sub-relativistic outflow has a velocity of $0.12c$. The data of AT2017gfo,
\citep{Pian2017,Kasliwal2017b,Troja2017,Tanvir2017} with the proper distance and extinction correction, are shown for further comparison.}
\hfill
\end{figure}

As a result of the sun site constraint, the optical/infrared observations of GRB 111005A were very rare.
Benefited from its very low redshift, the optical/infrared upper limits still impose interesting bounds on the macronova model. We focus on the double neutron star merger model since the neutron star-black hole mergers rate is widely expected to be much less frequent, i.e., about one order of magnitude lower than the double neutron star mergers \citep{LVC2010,LiX2017}. In Fig.\ref{fig:comparision}, we compare the $r, ~i, ~J, ~K$ band upper limits of GRB 111005A to the macronova emission predicted in the NSM-all model of \citet{2013ApJ...775..113T}. Note that the extinction of the host with $A_{\rm V}=2$ mag \citep{Michal2017} has been corrected. The sub-relativistic outflow has a rest mass of $0.01M_\odot$ and a velocity of $0.12c$ (where $c$ is the speed of the light). Interestingly, the predicted macronova emission are roughly consistent with the upper limits. We thus set a bound on the neutron-rich ejecta mass of GRB 111005A, i.e., $M_{\rm ej}\leq 0.01M_\odot$.

After submitting this work, the data of AT2017gfo \citep[e.g.][]{Coulter2017} have been released. In Fig.\ref{fig:comparision} we compare  AT2017gfo to the upper limits of GRB 111005A \citep[see also][]{Yue2017}. In all bands, the macronova emission associated GRB 111005A should be dimmer than AT2017gfo by a factor of $3-10$. Such a difference may be mainly caused by the large amount of r-process material ($M_{\rm ej}\sim 0.04\pm0.01M_\odot$) ejected from GRB 170817A. The presence of some lanthanides-free material in the directions surrounding the rotational axis of the remnant \citep{Pian2017,Kasliwal2017b} is needed to be explain the early bright multi-wavelength of AT201gfo. While for GRB 111005A, the tight constraints on the early optical emission may favor the  absence of the lanthanides-free material, indicating the central engine collapsed promptly (or one of the pre-merger object is a stellar mass black hole) and the disk wind outflow component did not play an important role \citep{Kasen2015}.

\begin{figure}[ht!]
\figurenum{2}\label{fig:Mej}
\centering
\includegraphics[angle=0,scale=0.6]{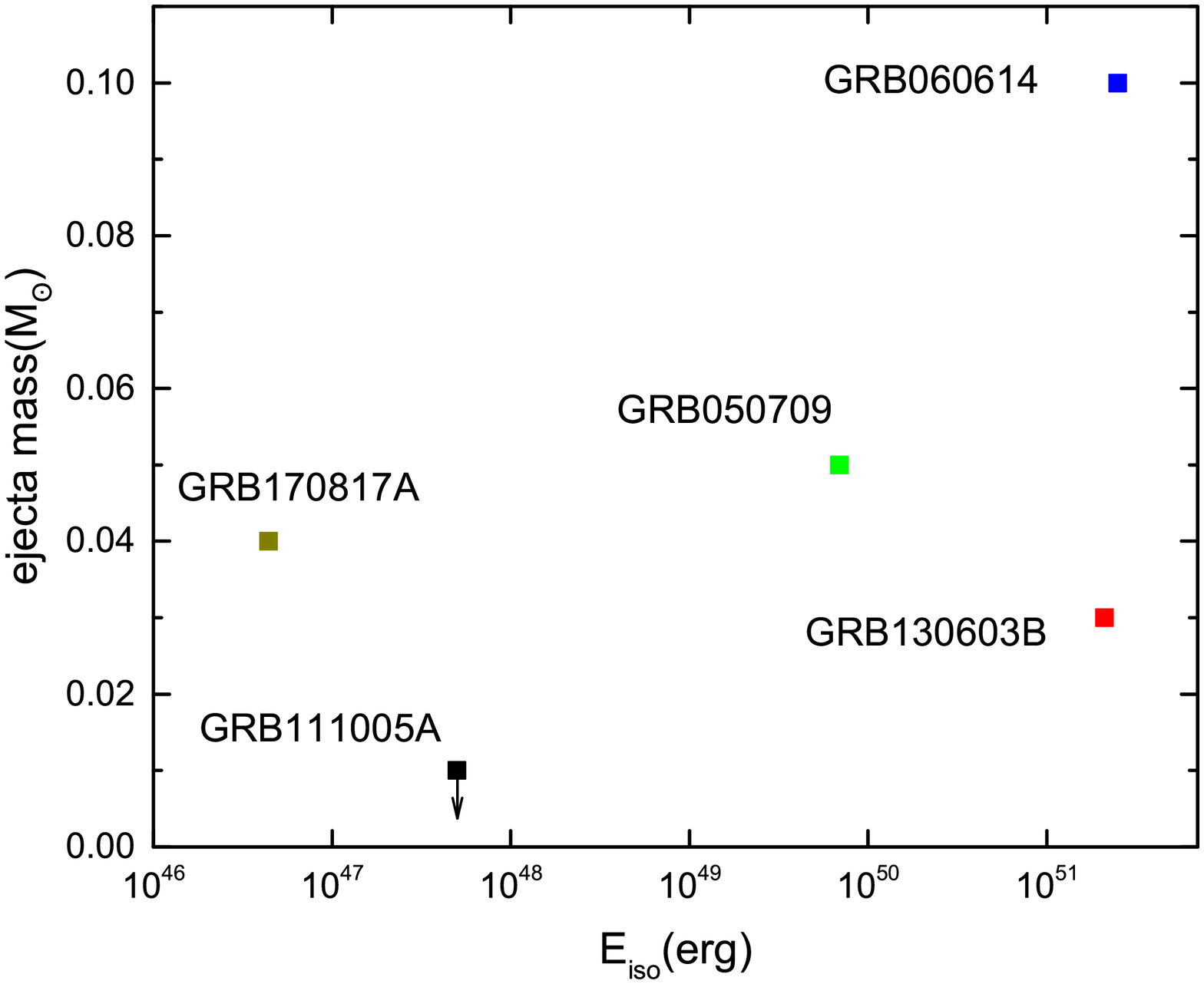}
\caption{The masses (or upper limit) of the neutron-rich ejecta of GRB 130603B, GRB 060614, GRB 050709, GRB 170817A and GRB 111005A. See the main text for the references.}
\hfill
\end{figure}

The numerical modeling of the
macronova signal of sGRB 130603B yields $M_{\rm ej}\sim 0.03 M_\odot$ \citep{Tanvir2013,Berger2013}. For GRB 060614 and GRB 050709, the macronova modeling gives a
$M_{\rm ej}\sim (0.1, ~0.05)~M_\odot$, respectively \citep{Yang2015,Jin2015,Jin2016}.
In Fig.\ref{fig:Mej} we summarize these results together with the upper limits for GRB 111005A.
Interestingly, \citet{Hotokezaka2015} showed that for current neutron star merger rate inferred from the short GRB data, each event should eject $\sim 10^{-2}-10^{-1}M_\odot$ r-process material to reproduce that observed in the Galaxy \citep[see also][]{WangH2017}. While the analysis of the r-process material in ultrafaint dwarf galaxies suggests $\sim 6\times 10^{-3}-4\times 10^{-2}M_\odot$ heavy elements in each neutron star merger \citep{Beniamini2016a,Beniamini2016b}. We therefore conclude that the ejecta masses of neutron star mergers are diverse and GRB 111005A may have launched relatively a small amount of sub-relativistic neutron-rich outflow (otherwise AT2017fgo is atypical).
Indeed the ejecta mass depends sensitively on the equation of state and on the mass ratio of the pre-merger compact objects \citep[e.g.,][]{Hotokezaka2013,Dietrich2015,Kawaguchi2015,Kyutoku2015}. In particular, for neutron star-black hole mergers, the black hole spin also plays an important role and  $M_{\rm ej}$ can be in a very wide range, from $\sim 0M_\odot$ to $\sim 0.2M_\odot$ (see e.g. Fig.1 of \citet{Shen2017} for a summary; please note that the data of GW170817 favor the equation of state models that predict a compact star \citep{LVC2017}, for which the neutron star-black hole mergers launch less massive ejecta than that believed before). One potential challenge for the neutron star-black hole merger model is the relatively low rate of such a kind of events \citep[usually it is lower than the neutron star merger rate by a factor of 10;][]{LVC2010}. As for double neutron star mergers, the mass ratios are expected to be much narrowly distributed, as observed in the Galaxy, and $M_{\rm ej}$ is mainly governed by the equation of state. One may thus expect a narrow distribution of $M_{\rm ej}$ as well. However, the disk wind as well as the neutrino-driven wind from the surface of the nascent hypermassive/supramassive neutron star formed in the merger can also enhance the neutron-rich outflow \citep{2017LRR....20....3M}. Therefore, a wide distribution of $M_{\rm ej}$ for double neutron star mergers is still possible. For a neutron star merger rate of ${\cal R}_{\rm nsm}\sim 10^{3}~{\rm Gpc^{-3}~yr^{-1}}$, as inferred from the successful detection of GW170817 in the second observational run of advanced LIGO \citep{LVC2017} and from the data of ``local" sGRBs \citep{Jin2017}\footnote{ The intrinsic sGRB rate (i.e., the neutron star merger rate) has been extensively investigated in the literature \citep[e.g.,][]{Ghirlanda2016,Coward2012,Wanderman2015}. Though these earlier approaches were based on the data of short GRBs at relatively high redshifts, the inferred neutron star merger rates are in the range of $\sim 10^{2}-10^{3}~{\rm Gpc^{-3}~yr^{-1}}$, reasonably consistent with that yielded from the local sGRB data and from GW170817.}, a reasonably large neutron star merger sample will be available in the near future and the dedicated macronova/kilonova observations and modeling will yield a reliable distribution of $M_{\rm ej}$, with which the double neutron star merger origin possibility of GRB 111005A will be directly tested.

\yzf{}

\section{The rate density of lsGRBs and the gravitational wave detection prospect}
GRB 111005A has a $E_{\rm iso}\sim 10^{47}$ ergs and a $E_{\rm p}<15$ keV, which could well be an off-beam (if the GRB outflow is uniform) or off-axis (if the GRB outflow is structured) short GRB \citep[As shown in][some bright sGRBs could reproduce the characters of the prompt emission of GRB 111005A if viewed off-beam]{Yue2017}. GRB 060505 has a $E_{\rm iso}\sim 10^{49}$ erg and a $T_{90}\sim 4$ s but a hard spectrum \citep{Ofek2007}.
Therefore, the off-beam/off-axis GRB scenario is disfavored. As for GRB 060614, it is so bright/long that is very unlikely to be an off-beam/axis event. We thus conclude that not all the lsGRBs as the off-beam/axis sGRBs and it is thus necessary to pin down their progenitors. For such a purpose, the rate of lsGRBs is needed. The lack of jet half-opening angle information of GRB 060505 and GRB 111005A does not hamper since in this work we are keen on the lsGRB/GW association events, only for which the progenitors of lsGRBs can be directly revealed. ¡¡

Inspired by the method utilized in \citet{2016PhRvX...6d1015A} to determine the binary black hole merger rates, we estimate the lsGRB rate in the same way. The main point of the procedure is to relate the rate and the observation with $\Lambda ={\cal R}\left\langle {VT} \right\rangle$, where $\Lambda$ is the poisson mean number of astrophysical trigger (our current approach is much simpler than that for GWs, since the three lsGRBs are well identified as gamma-ray burst and we do not need to consider terrestrial trigger), and $\left\langle {VT} \right\rangle$ is the population-averaged sensitive space-time volume of the search. Generally, $\left\langle {VT} \right\rangle$ can be calculated with \citep{Liang2007}
\begin{equation}\label{VT}
\left\langle {VT} \right\rangle  = \frac{{\Omega T}}{{4\pi }}\int_{{L_{\min }}}^{{L_{\max }}} {\Phi \left( L \right)dL\int_0^{{z_{\max }}} {\frac{1}{{1 + z}}\frac{{d{V_c}\left( z \right)}}{{dz}}dz} }
\end{equation}
where $\Omega$ is the field of view of instrument, $\Phi \left( L \right)$ is the luminosity function of lsGRBs, and $z_{\rm max}$ corresponds to the maximum detection range for a burst with luminosity $L$ and is determined by the instrument flux threshold. Having the fact that the luminosity function can not be well constrained since there are just three identified lsGRBs so far, the result of the integration in Eq.(\ref{VT}) is model dependent. To avoid the large uncertainties in the luminosity function, we calculate the rate based upon the properties of individual events (event based), i.e., GRB 111005A, GRB 060505 and GRB 060614 are treated as three distinct classes that together stands for the whole population of lsGRBs, and their $\left\langle {VT} \right\rangle$ are obtained independently by
\begin{equation}\label{VT}
	\left\langle {VT} \right\rangle_{\rm i}  = \frac{\Omega T}{4\pi }\int_0^{{z_{\rm \max,i }}} {\frac{1}{{1 + z}}\frac{{d{V_c}\left( z \right)}}{{dz}}dz}
\end{equation}
(see the following discussion for the choice of $z_{\rm \max,i }$ for each burst). The total event rate is then the sum of the individual rates derived from each  $\left\langle {VT} \right\rangle$. Such an approach is different from that used in previous GRB rate estimate \citep[e.g.,][]{Liang2007}.

Assuming a poisson fluctuation on the observed number of event (in our case the observed number is 1 for each class), the likelihood for the rate ${\cal R}$ of a given class is
\begin{equation}\label{llh}
{\cal L}\left( {1|{\cal R}} \right) = \Lambda {e^{ - \Lambda }} ={\cal R}\left\langle {VT} \right\rangle \exp \left[ { - {\cal R}\left\langle {VT} \right\rangle } \right]
\end{equation}
The posterior over ${\cal R}$ is then obtained by multiplying the likelihood with a prior $P({\cal R})$ and normalized over the possible range of ${\cal R}$. Two kinds of function are chosen as our prior: an Uniform distribution of ${\cal R}$ and a Poisson Jeffreys prior which proportional to $1/\sqrt {\cal R}$.

\begin{figure}[ht!]
\figurenum{3}\label{fig:density}
\centering
\includegraphics[angle=0,scale=1.0]{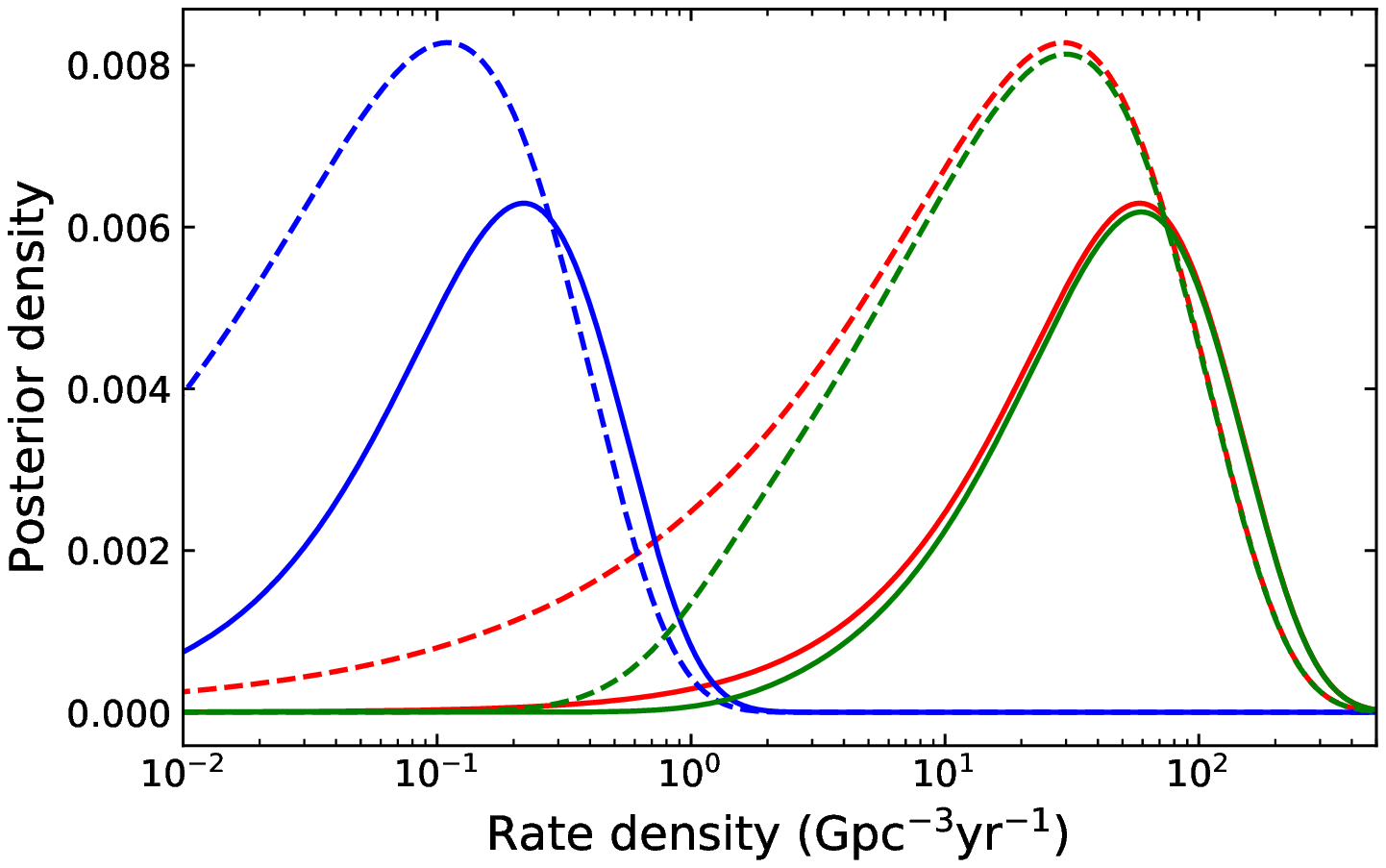}
\caption{The posterior distribution of GRB 111005A class (red), GRB 060614 (or GRB 060505) class (blue) and the total lsGRB population (green). The solid lines and the dashed lines are derived from Uniform prior and Poisson Jeffreys prior respectively. The probability density of GRB 060614 (or GRB 060505) class has been rescaled for comparison.}
\hfill
\end{figure}

We first calculate the rate for GRB 111005A class GRBs following the procedure described above. We collect the 1-second peak energy flux in $15-150$ keV band of GRB 111005A from Swift/BAT GRB Catalog \citep{2016ApJ...829....7L}. At the redshift of 0.0133, the luminosity in this band is $2.8 \times 10^{46}~{\rm erg~s^{-1}}$. As the Swift/BAT threshold is $\sim 10^{-8} ~{\rm erg~s^{-1}~cm^{-2}}$, such a low luminosity event can only be seen within $z=0.035$, implying a very small search volume and thus a very high astrophysical rate density. By applying eq.(\ref{llh}), considering Swift has a field of view $\sim 1.4$ sr and 11 years of observation \citep{Gehrels2004}, a rate for the GRB 111005A class is found to be ${\cal R}_{\rm GRB 111005A}=58^{+219}_{-38} ~{\rm Gpc^{-3}~yr^{-1}}$ (using the Uniform prior, the errors are reported in 90 percent confidence level), or $29^{+199}_{-18}~{\rm Gpc^{-3}~yr^{-1}}$ (using the Poisson Jeffreys prior), which is about one order of magnitude higher than the sGRB rate \citep{Wanderman2015}. The posterior distribution are shown in Figure \ref{fig:density}. The distribution is affected by the prior, and this can be understood by the fact that the Uniform prior extends the probability density to infinity and hence may overestimate the rate at high end, while the Poisson Jeffreys prior has a steep decay shape and thus underestimate the rate at high end.
The detection rate of GRB 111005A-like event by a full-sky monitor with a sensitivity comparable to {\it Swift}, for example the proposing GECAM \citep[the Gravitational Wave Electromagnetic Counterpart Monitor;][]{Xiong2017}, would thus be (for the Uniform prior)
\begin{equation}
R_{\rm GRB 111005A} = {\cal R}_{\rm GRB111005A}V_{(z\leq 0.035)}\approx 0.82^{+3.06}_{-0.53}~{\rm yr^{-1}},
\end{equation}
where $V_{(z\leq 0.035)}$ is the comoving volume for an assumed detection horizon of $z=0.035$. Such a possible detection rate of lsGRB/GW association is already comparable to that of the bight sGRB/GW association though the latter can be detected up to a distance of $\sim 400$ Mpc \citep[e.g.,][]{2016ApJ...827...75L}.

The ${\cal R}_{\rm GRB 111005A}\sim 60~{\rm Gpc^{-3}~yr^{-1}}$ (note that no jet half-opening angle correction has been made) matches the local neutron star merger rate ${\cal R}_{\rm nsm}\sim 10^{3}~{\rm yr^{-1}}$ if the jet half-opening angle of GRB 111005A-like events is typically $\sim 0.4$ rad, which is significantly larger than that for bright sGRBs (i.e., $\theta_{\rm j}\sim 0.1$ rad), and thus favors the off-axis structured sGRB jet model or even the cocoon radiation scenario \citep{Jin2017,Lazzati2017}.

Different from GRB 111005A, the events of GRB060614 and GRB060505 are more luminous and can be detected by {\it Swift} up to $z\sim 0.63$ and $\sim 1.4$ respectively. However, the reliable identification of lsGRBs at relatively high redshifts is quite a challenging. So far, the furthest lsGRB candidate is XRF 040701 at a redshift of 0.21 \citep{Soderberg2005}. Therefore the corresponding $\left\langle {VT} \right\rangle$ are limited by the identification probability, rather than their luminosities. We assume a ``valid" identification horizon for lsGRBs as  $z\sim 0.25$ and present the resulting posterior distribution for GRB 060505 class and GRB 060614 class in Figure \ref{fig:density} (they share the same distribution since their identification horizons are assumed to be the same), and the inferred rate is ${\cal R}_{\rm GRB060505}\sim 0.22^{+0.82}_{-0.14}~{\rm Gpc^{-3}~yr^{-1}}$ (the Uniform prior) or $0.11^{+0.75}_{-0.07}~{\rm Gpc^{-3}~yr^{-1}}$ (the Poisson Jeffreys prior). ${\cal R}_{\rm GRB060505}$ is about two orders of magnitudes lower than ${\cal R}_{\rm GRB 111005A}$, implying that GRB 111005A may be different from the other two, as already speculated in the previous paragraph. Finally, the posterior distribution of the total rate density of lsGRBs is calculated by convoluting the posterior distributions of the three classes, and the rates for uniform and Jeffreys prior are $59^{+220}_{-36} ~{\rm Gpc^{-3}~yr^{-1}}$ and $30^{+205}_{-18} ~{\rm Gpc^{-3}~yr^{-1}}$, respectively. As expected, the total rate density is dominated by the GRB 111005A class (see Figure \ref{fig:density}).


\section{Summary} \label{sec:discussion}
GRB 111005A is the closest lsGRB reliably identified so far. This burst, with a $E_{\rm iso}\sim 10^{47}$ ergs and a $E_{\rm p}<15$ keV, could well be an off-beam or off-axis short GRB; while for GRB 060505 and GRB 060614, such a possibility is strongly disfavored.
The infrared/optical upper limits of GRB 111005A, though rare, still impose tight constraint on the mass of the neutron-rich outflow expected in the neutron star merger scenario. The inferred bound (i.e., $M_{\rm ej} \leq 0.01~M_{\rm ej}$) is significantly smaller than that found in macronova modeling of GRB 130603B, GRB 060614 and GRB 050709. A neutron star-black hole merger can just eject a tiny amount of neutron-rich material. Their merger rate, however, is usually expected to be significantly lower than the neutron star mergers hence such a kind of events should be less frequent. On the other hand, a wide distribution of $M_{\rm ej}$ for double neutron star mergers is still possible and reliable measurements are expected in the advanced LIGO/Virgo era, with which the binary neutron star merger origin of GRB 111005A will be directly tested. The successful identification of three nearby lsGRB among {\it Swift} events suggests a high rate of $59^{+220}_{-36} ~{\rm Gpc^{-3}~yr^{-1}}$ (for the Uniform prior) and $30^{+205}_{-18} ~{\rm Gpc^{-3}~yr^{-1}}$ (the Poisson Jeffreys prior), note that no jet opening-angle correction is made since we are keen on the lsGRB/GW association. If most lsGRBs were indeed from neutron star mergers, the prospect of establishing the lsGRB/GW association is very promising. The successful detection will thus close the debate on the physical origin of lsGRBs. The collaboration of a full-sky monitor with a sensitivity of {\it Swift} is crucial for such a purpose. Interestingly, the proposing GECAM is such an instrument.
In 2020s if no significant GW signals coincident with the lsGRBs have been recorded, with the help of advanced optical/infrared telescopes, the environmental conditions at the position of some very-nearby long-short GRBs will be well examined. If environmental condition similar to that of GRB 111005A can be identified for a good fraction of long-short events, a novel kind of merger origin rather than a new type of collapsar model will be favored. Such mergers could involve the black hole-white dwarf \citep{Fryer1999} or/and the black hole-helium star \citep{Fryer1998} binaries.

In the final stage of the preparation of this work, \citet{Dado2017} appeared and the authors suggested a neutron star phase transition model for the lsGRBs. If correct, no lsGRB/GW association is expected in the era of the second generation gravitational wave detectors, either.

\section*{Acknowledgments}
We thank the anonymous referee for helpful suggestions. This work was supported in part by 973 Programme of China (No. 2014CB845800), by NSFC under grants 11525313
(the National Natural Fund for Distinguished Young Scholars), 11273063, 11433009 and 11763003, by the Chinese Academy of Sciencesvia the Strategic Priority Research Program (No. XDB09000000) and the External Cooperation Program of BIC (No.114332KYSB20160007). F.-W.Z. also acknowledges the support by the Guangxi Natural Science
Foundation (No. 2017GXNSFAA198094).

\clearpage

\end{document}